\documentclass[12pt]{article}
\usepackage{amsmath}

\usepackage{amssymb}

\def\hybrid{\topmargin 0pt      \oddsidemargin 0pt
        \headheight 0pt \headsep 0pt
        \voffset=-0.5cm
        \textwidth 6.25in       
        \textheight 9.5in       
        \marginparwidth 0.0in
        \parskip 5pt plus 1pt   \jot = 1.5ex}
\catcode`\@=11
\def\marginnote#1{}

\newcount\hour
\newcount\minute
\newtoks\amorpm
\hour=\time\divide\hour by60 \minute=\time{\multiply\hour by60
\global\advance\minute by-\hour}
\edef\standardtime{{\ifnum\hour<12 \global\amorpm={am}%
        \else\global\amorpm={pm}\advance\hour by-12 \fi
        \ifnum\hour=0 \hour=12 \fi
        \number\hour:\ifnum\minute<10 0\fi\number\minute\the\amorpm}}
\edef\militarytime{\number\hour:\ifnum\minute<10 0\fi\number\minute}

\def\draftlabel#1{{\@bsphack\if@filesw {\let\thepage\relax
   \xdef\@gtempa{\write\@auxout{\string
      \newlabel{#1}{{\@currentlabel}{\thepage}}}}}\@gtempa
   \if@nobreak \ifvmode\nobreak\fi\fi\fi\@esphack}
        \gdef\@eqnlabel{#1}}
\def\@eqnlabel{}
\def\@vacuum{}
\def\draftmarginnote#1{\marginpar{\raggedright\scriptsize\tt#1}}
\def\draftlabel#1{{\@bsphack\if@filesw {\let\thepage\relax
   \xdef\@gtempa{\write\@auxout{\string
      \newlabel{#1}{{\@currentlabel}{\thepage}}}}}\@gtempa
   \if@nobreak \ifvmode\nobreak\fi\fi\fi\@esphack}
        \gdef\@eqnlabel{#1}}
\def\@eqnlabel{}
\def\@vacuum{}
\def\draftmarginnote#1{\marginpar{\raggedright\scriptsize\tt#1}}

\def\draft{\oddsidemargin -.5truein
        \def\@oddfoot{\sl preliminary draft \hfil
        \rm\thepage\hfil\sl\today\quad\militarytime}
        \let\@evenfoot\@oddfoot \overfullrule 3pt
        \let\label=\draftlabel
        \let\marginnote=\draftmarginnote
   \def\@eqnnum{(\theequation)\rlap{\kern\marginparsep\tt\@eqnlabel}%
\global\let\@eqnlabel\@vacuum}  }

\def\numberbysection{\@addtoreset{equation}{section}
        \def\theequation{\thesection.\arabic{equation}}}

\def\underline#1{\relax\ifmmode\@@underline#1\else
        $\@@underline{\hbox{#1}}$\relax\fi}

\def\titlepage{\@restonecolfalse\if@twocolumn\@restonecoltrue\onecolumn
     \else \newpage \fi \thispagestyle{empty}\c@page\z@
        \def\thefootnote{\fnsymbol{footnote}} }

\def\endtitlepage{\if@restonecol\twocolumn \else  \fi
        \def\thefootnote{\arabic{footnote}}
        \setcounter{footnote}{0}}  

\numberbysection \hybrid

\newcommand{\tr}{{\rm tr}}

\newcommand{\la}{\lambda}

\newcommand{\al}{\alpha}
\newcommand{\be}{\beta}
\newcommand{\ga}{\gamma}
\newcommand{\om}{\omega}

\def\beq{\begin{equation}}
\def\eeq{\end{equation}}
\def\p{\partial}

\newtheorem{theor}{Theorem}

\def\res{\mathop{\hbox{Res}}\limits}

\begin{document}

\setcounter{page}{1}

\date{}
\date{}
\vspace{50mm}

\begin{flushright}
 ITEP-TH-26/15\\
\end{flushright}
\vspace{0mm}

\begin{center}
\vspace{2mm}
{\LARGE{Trigonometric version of quantum-classical}}
 \\ \vspace{4mm}
 {\LARGE{duality in integrable systems}}
\\
\vspace{12mm} {\large { M. Beketov}\,$^{\sharp}$\ \ \  { A.
Liashyk}\,$^{\flat\, \ddag}$\ \ \  { A. Zabrodin}\,$^{\flat\, \dag\,
\natural}$\ \ \
 { A. Zotov}\,$^{\diamondsuit \natural\, \sharp}$}\\ \vspace{5mm}
 \vspace{3mm} $^\sharp$ - {\small{ MIPT, Inststitutskii per.  9, 141700, Dolgoprudny,
 Moscow region, Russia}}\\
 \vspace{2mm} $^\flat$ - {\small{ National Research University Higher School of
Economics,  Myasnitskaya str. 20,\\ 101000, Moscow, Russia}}\\
 \vspace{2mm} $^\ddag$ - {\small{ BITP, Metrolohichna str. 14-b, 03680, Kiev, Ukraine}}\\
 \vspace{2mm}$^\natural$ - {\small{ 
 ITEP, Bolshaya Cheremushkinskaya str. 25, 117218,  Moscow, Russia}}\\
 \vspace{2mm} $^\dag$ - {\small{ Institute of Biochemical Physics,
Kosygina str. 4, 119991, Moscow, Russia}}\\
  \vspace{2mm} $^\diamondsuit$ - {\small{ Steklov Mathematical Institute, RAS, Gubkina str. 8, 119991, Moscow,
  Russia}}
\end{center}

\begin{center}\footnotesize{{\rm E-mails:}{\rm\ \
 beketov@phystech.edu,\ a.liashyk@gmail.com,\ zabrodin@itep.ru,\  zotov@mi.ras.ru}}\end{center}

 \begin{abstract}
We extend the quantum-classical duality to the
trigonometric (hyperbolic) case. The duality establishes an explicit
relationship between the classical $N$-body tri\-go\-no\-met\-ric
Ruijsenaars-Schneider model and the inhomogeneous twisted
XXZ spin chain on $N$ sites. Similarly to the
rational version, the spin chain data fixes a certain Lagrangian
submanifold in the phase space of the classical integrable system.
The inhomogeneity parameters are equal to the coordinates of
particles while the velocities of classical particles are
proportional to the eigenvalues of the spin chain Hamiltonians (residues
of the properly normalized transfer matrix). In the rational version of
the duality, the action variables of the Ruijsenaars-Schneider model are
equal to the twist parameters with some multiplicities defined by
quantum (occupation) numbers. In contrast to the rational version, in
the trigonometric case there is a splitting of the spectrum of action
variables (eigenvalues of the classical Lax matrix). The limit
corresponding to the classical Calogero-Sutherland system and
quantum trigonometric Gaudin model is also described as well as the XX
limit to free fermions.

 \end{abstract}

\newpage


\section{Introduction}
\setcounter{equation}{0}

\noindent
The quantum-classical (QC) duality (correspondence) is an explicit
relation between quantum and classical integrable systems of
different types. This phenomenon was first
observed in \cite{Givental} for the classical Toda chain.
A similar observation was made in \cite{MTV1} for the classical
Calogero system and quantum Gaudin model. The classical action variables
were assumed to be equal to zero. The case of arbitrary set of action
variables  was described in \cite{zabrodin2} using the relation of both
models to the KP hierarchy \cite{zabrodin3}. In a similar way, the QC
duality
 between the classical Ruijsenaars-Schneider (RS) model
and the quantum twisted spin chain was proposed in
\cite{zabrodin1,zabrodin1a}. The final version and a direct proof of
this relation was presented in \cite{GZZ} via the nested
Bethe anzats. Later the duality was extended to the correspondence
\cite{TsuboiZZ}: it was shown that the RS model
is related not to a single quantum model but to a family of
supersymmetric spin chains. We do not discuss the supersymmetric case
in this paper.

Let us briefly recall the result of \cite{GZZ}.
Consider the Lax matrix of the classical $N$-body
RS model\footnote{In (\ref{tt01}) the sets
of variables $\{{\dot q}_j\}$ and $\{q_j\}$ are velocities and
coordinates of particles respectively, $\nu$ is the coupling
constant and $\eta$ is the relativistic deformation parameter
(inverse light speed).} \cite{Ruijs1}
 \beq\label{tt01}
 L^{\hbox{\tiny{RS}}}_{ij}=\frac{\nu\, {\dot q}_j}{q_i-q_j+\eta\nu}\,,\ \ \ i,j=1\,,...\,,N
  \eeq
and quantum transfer matrix of the $GL(n)$ inhomogeneous
(generalized) twisted XXX spin chain on $N$ sites\footnote{In
(\ref{tt02}) $\{\hat{H}^{\hbox{\tiny{XXX}}}_j\}$ are the quantum
(non-local) Hamiltonians , $\{z_j\}$ are  inhomogeneity parameters and
$V=\hbox{diag}(V_1,...,V_n)$ is the twist matrix.}
 \beq\label{tt02}
\hat{T}^{\hbox{\tiny{XXX}}}(z)= \mbox{tr} \, V
+\sum_{j=1}^{N}\frac{\hat{H}^{\hbox{\tiny{XXX}}}_j}{z-z_j}\,.
  \eeq
In the framework of the algebraic nested Bethe ansatz the spectrum
${H}^{\hbox{\tiny{XXX}}}_j$ of the Hamiltonians
$\{\hat{H}^{\hbox{\tiny{XXX}}}_j\}$ is constructed in terms of the
Bethe roots $\Bigl \{\{\mu^1_i\}_{N_1}, \ldots ,
\{\mu^{n-1}_i\}_{N_{n-1}}\Bigr \}$ which are solution of the system
of Bethe equations. Here  $N_{c}$ is the number of Bethe roots at
the $c$-th level of the nested Bethe ansatz.

Substitute
 \beq\label{tt03}
\nu\eta=\hbar\,,
  \eeq
 \beq\label{tt031}
q_j=z_j\,,\ \ \  j=1...N
  \eeq
and
 \beq\label{tt04}
{\dot q}_j=\frac{\eta}{\hbar}\, H^{\hbox{\tiny{XXX}}}_j \Bigl
(\{q_i\}_N; \{\mu^1_i\}_{N_1}, \ldots ,
\{\mu^{n-1}_i\}_{N_{n-1}}\Bigr )\,,\ \ \ j=1\,,...\,,N\,,
  \eeq
  where  $\{\mu_i^a\}$ is any solution of the Bethe equations. Then
  the spectrum of the classical Lax matrix (\ref{tt01}) is given by
  the twist parameters:
 \beq\label{tt05}
 \begin{array}{c}
 \big(\underbrace{V_1\,,\ldots\,,V_1}_{N-N_1}\,,\underbrace{V_2\,,\ldots\,,V_2}_{N_1-N_2}\,,\ldots\,,
 \underbrace{V_{n\!-\!1}\,,\ldots\,,V_{n\!-\!1}}_{N_{n\!-\!2}-N_{n\!-\!1}}\,,
 \underbrace{V_n\,,\ldots\,,V_n}_{N_{n\!-\!1}}\big ).
\end{array}
  \eeq
The multiplicities are defined by the quantum numbers $N_c$.

Let us also mention that the QC correspondence appeared also in the
framework of gauge theory dualities \cite{NRS,GK13,Gorsky}. Another
relation between classical Lax matrices and quantum $R$-matrices
related to spin chains can be found in \cite{LOZ}.

\vskip3mm

The purpose of this paper is the trigonometric version of the QC duality.
We prove an analogue of statement (\ref{tt05}) for the
trigonometric (hyperbolic) RS model and the XXZ twisted inhomogeneous
spin chain. We show that in contrast to the rational
version, the degeneration of the spectrum of
action variables (eigenvalues of the classical Lax matrix)
disappears. The identification
 \beq\label{tt06}
\eta\nu=\hbar
  \eeq
and
 \beq\label{tt07}
 {\dot q}_j=
 \frac{\eta}{\sinh \hbar}\,H_{j}^{\hbox{\tiny{XXZ}}}
  \eeq
leads to the following  eigenvalues of the classical RS Lax matrix
(to be compared with (\ref{tt05}) for the rational case):
 \beq\label{tt08}
 \begin{array}{c}
 \displaystyle{\Big\{\underbrace{e^{-(N-N_1-1)\hbar}
 \,V_1,\ldots,e^{(N-N_1-1)\hbar}\,V_1}_{N-N_1}\,, \,
 \underbrace{e^{-(N_1-N_2-1)\hbar}\,V_2,\ldots,e^{(N_1-N_2-1)
 \hbar}\,V_2}_{N_1-N_2}\,,\ldots\,,
 }
\end{array}
  \eeq
 $$
 \displaystyle{
 \ldots\,,\underbrace{e^{-(N_{n-2}-N_{n-1}-1)\hbar}\,V_{n-1},
 \ldots,
 e^{(N_{n-2}-N_{n-1}-1)\hbar}\,V_{n-1}}_{N_{n-2}-N_{n-1}}\,,\,
 \underbrace{e^{-(N_{n-1}-1)\hbar}\,V_n,\ldots,e^{(N_{n-1}-1)\hbar}\,V_n}_{N_{n-1}}\Big\}}
 $$
The eigenvalues of the Lax matrix form ``strings'' centered at
the twist parameters $V_a$.

{\small
\noindent {\bf Acknowledgments.}
The work of A.L. was supported in part
by the Ukrainian-Russian (NASU-RFBR) project 01-01-14.
The work of A. Zabrodin was supported in part by
RFBR grant 14-02-00627, by joint grants 15-52-50041-YaF, 14-01-90405-Ukr and by grant
 for support of scientific schools NSh-1500.2014.2.
The work of A. Zotov was partially supported by RFBR  project
15-31-20484 mol$\_$a$\_$ved and by the D. Zimin's fund "Dynasty".
The article was prepared within the framework of a subsidy granted
to the HSE by the Government
 of the Russian Federation for the implementation of the Global Competitiveness Program.
}

\section{Trigonometric Ruijsenaars-Schneider model}
\setcounter{equation}{0}
In this paper we use the following Lax matrix of the trigonometric
$N$-particle RS model:
 \beq\label{t01}
 \displaystyle{
L^{\hbox{\tiny{RS}}}_{ij}=\frac{\sinh(\eta\nu)}{\sinh(q_i-q_j+\eta\nu)}\,e^{\eta
p_j}\, \prod\limits_{k\neq
j}^N\frac{\sinh(q_j-q_k-\eta\nu)}{\sinh(q_j-q_k)}\,,\quad
i,j=1\,,...\,,N.
 }
  \eeq
The Hamiltonian is defined as
 \beq\label{t02}
 H^{\hbox{\tiny{RS}}}=\tr L^{\hbox{\tiny{RS}}}=
 \sum\limits_{j=1}^N e^{\eta p_j}\prod\limits_{k\neq
j}^N\frac{\sinh(q_j-q_k-\eta\nu)}{\sinh(q_j-q_k)}\,.
  \eeq
For the velocities we have
 \beq\label{t03}
{\dot q}_j ={\frac{\p H}{\p p_i}}^{\hbox{\tiny{RS}}}= \eta\, e^{\eta
p_j}\prod\limits_{k\neq
j}^N\frac{\sinh(q_j-q_k-\eta\nu)}{\sinh(q_j-q_k)}\,.
  \eeq
Therefore, in terms of velocities the Lax matrix has the form
 \beq\label{t04}
 \displaystyle{
L^{\hbox{\tiny{RS}}}_{ij}=\frac{\sinh(\eta\nu)}{\sinh(q_i-q_j+\eta\nu)}\,\eta^{-1}
{\dot q}_j
 }
  \eeq
or
 \beq\label{t05-1}
 \begin{array}{l}
 \displaystyle{
L^{\hbox{\tiny{RS}}}_{ij}=\eta^{-1}\,C_{ik}\,\dot{Q}_{kj}\,,~Q_{ij}=\delta_{ij}\,q_j\,,}
 \\ \ \\
 \displaystyle{
C_{ij}=\frac{\sinh(\eta\nu)}{\sinh(q_i-q_j+\eta\nu)}\,,
 \quad
i,j=1\,,...\,,N.}
\end{array}
  \eeq
Here $||C_{ij}||$ is the trigonometric Cauchy matrix.

 It is important
for our purpose that the classical Lax matrix (\ref{t01}) admits
the following factorization (see
\cite{Hasegawa,AASZ}):
 \beq\label{t05}
 \displaystyle{
L^{\hbox{\tiny{RS}}}=D\, {\tilde V}^{-1}(\epsilon) {\tilde
V}(\epsilon-\eta\nu) D^{-1} e^{\eta P}\,,
 }
  \eeq

\noindent where $P=\hbox{diag}(p_1,...,p_N)$, ${\tilde V}$ is the
(trigonometric) Vandermonde type matrix
 \beq\label{t06}
 \displaystyle{
{\tilde V}_{ij}(\epsilon)=\exp\left( (2i-1-N)(q_j+\epsilon) \right)
 }
  \eeq
and
 \beq\label{t07}
 \displaystyle{
D_{ij}=\delta_{ij}\prod\limits_{k\neq j}^N \sinh(q_j-q_k)\,.
 }
  \eeq
The (spectral) parameter $\epsilon$ is fictitious -- it does not
enter the final answer. Notice that
 \beq\label{t08}
 \displaystyle{
{\tilde V}(\epsilon-\eta\nu)=S(\eta\nu) {\tilde V}(\epsilon)\,,
 }
  \eeq
where $S$ is the following diagonal matrix:
 \beq\label{t09}
 \displaystyle{
S_{ij}(\zeta)=\delta_{ij}\exp\left( -(2i-1-N)\zeta \right)\,.
 }
  \eeq
It follows from (\ref{t05}) and (\ref{t08}) that the eigenvalues of
the Lax matrix (\ref{t01}) become very simple on the Lagrangian
submanifold $P=0$ (i.e. $p_k=0$ for all $k=1,...,N$). The spectrum
of (\ref{t01}) is then given by the elements of matrix $S(\eta\nu)$:
 \beq\label{t10}
 \displaystyle{
{\rm
Spec}(L^{\hbox{\tiny{RS}}}\left.\right|_{P=0})=\left\{e^{-(N-1)\eta\nu},
e^{-(N-3)\eta\nu},...,e^{(N-3)\eta\nu},e^{(N-1)\eta\nu}\right\}\,.
 }
  \eeq

The equations of motion $\displaystyle{
{\dot p}_j =-{\frac{\p H}{\p q_i}}^{\hbox{\tiny{RS}}}}$ of the RS model
admit the Lax representation
$
\dot L^{\hbox{\tiny{RS}}}=[B^{\hbox{\tiny{RS}}}, \, L^{\hbox{\tiny{RS}}}]
$,
where
$$
B^{\hbox{\tiny{RS}}}_{jk}=\Bigl (
\sum_{l\neq j}\dot q_l \coth q_{jl} -\sum_l \dot q_l \coth (q_{jl}+\eta \nu)
\Bigr ) \delta_{jk} +\frac{1-\delta_{jk}}{\sinh q_{jk}},\quad q_{jk}\equiv q_j-q_k.
$$
Explicitly, the equations of motion
read
\beq\label{em1}
\ddot q_j= -\sum_{k\neq j}\frac{2\dot q_j \dot q_k \sinh ^2 (\eta \nu )
\cosh q_{jk}}{\sinh (q_{jk}-\eta \nu)\sinh q_{jk}\sinh (q_{jk}+\eta \nu)}.
\eeq

Two special cases of the trigonometric RS model are $\eta \nu =\pm \infty$ and
$\eta \nu =i\pi /2$. In the former case the equations of motion simplify to
\beq\label{em2}
\eta \nu =\pm \infty : \quad \ddot q_j=2\sum_{k\neq j}\dot q_j \dot q_k
\coth q_{jk}.
\eeq
In the latter case  they are:
\beq\label{em3}
\eta \nu = \frac{i\pi}{2}: \quad \ddot q_j =4\sum_{k\neq j}
\frac{\dot q_j \dot q_k}{\sinh (2q_{jk})}.
\eeq

\section{Inhomogeneous ${\mathcal U}_q(\hat{\mathfrak{gl}}_n)$ spin chain}
\setcounter{equation}{0}

The algebraic structure of the Heisenberg XXZ spin chain is based on the
quantum affine algebra ${\mathcal U}_q(\hat{\mathfrak{gl}}_n)$
\cite{Jimbo,FRT} (see also \cite{BR}).
The model is defined by the following quantum $R$-matrix:
 \beq\label{t20}
  \begin{array}{c}
 \displaystyle{
R_{12}(z)=\frac{\sinh (z+\hbar)}{\sinh z}\sum\limits_{a=1}^n
{\rm e}_{aa}\otimes {\rm e}_{aa}
+\sum\limits_{1\leq a\neq b\leq n} {\rm e}_{aa}\otimes {\rm e}_{bb}
 }
 \\
  \displaystyle{
+ \, \frac{\sinh \hbar}{\sinh z}\sum\limits_{1\leq
a\neq b\leq n}  e^{{\hbox{\tiny{sign}}}(b-a)z}\,
{\rm e}_{ab}\otimes {\rm e}_{ba}\,.
 }
\end{array}
  \eeq
  where $z$ is the spectral parameter, $\hbar$ is the anisotropy
  parameter and ${\rm e}_{ab}$ denotes the $n\! \times \! n$ matrix
  with 1 in the position $(a,b)$ and 0 otherwise.

The transfer matrix of the twisted inhomogeneous Heisenberg XXZ
model on $N$ sites is given by
 \beq\label{t22}
 \hat T^{\hbox{\tiny{XXZ}}}(z)=\,
 \tr_0\, \Bigl [V_0\,R_{01}(z-q_1)\,...\,R_{0N}(z-q_N)\Bigr ]\,,
  \eeq
where the diagonal twist matrix
 \beq\label{t23}
 V={\rm diag}(V_1,V_2, ..., V_n)\,.
  \eeq
  acts in the auxiliary $n$-dimensional vector space labeled by 0.
  We assume that the parameters $q_k$
  are in general position, i.e. $q_j \neq q_k$ and $q_j\neq q_k\pm \hbar$ for $j\neq k$.
  It follows from the Yang-Baxter equation for the $R$-matrix that the
  transfer matrices commute for different values of the spectral parameter:
  $[\hat T^{\hbox{\tiny{XXZ}}}(z), \, \hat T^{\hbox{\tiny{XXZ}}}(z')]=0$.

The nested Bethe ansatz gives the following result for
eigenvalues of the transfer matrix (\ref{t22}):
\beq\label{t24}
\begin{array}{l}\displaystyle{
  T^{\hbox{\tiny{XXZ}}}(z)=V_1 \prod_{k=1}^N
\frac{\sinh(z-q_k+\hbar )}{\sinh(z-q_k)}
\prod\limits_{\ga=1}^{\,N_{1}}\frac{\sinh(z-\mu_\ga^{1}
-\hbar)}{\sinh(z-\mu_\ga^{1})}}
\\ \\
\phantom{aaaaaaaa} +\, \displaystyle{
\sum\limits_{b=2}^n V_b \,
\prod\limits_{\ga=1}^{N_{b\!-\!1}}\frac{\sinh(z-\mu_\ga^{b\!-\!1}
+\hbar)}{\sinh(z-\mu_\ga^{b\!-\!1})}
\prod\limits_{\ga=1}^{\,N_{b}}\frac{\sinh(z-\mu_\ga^{b}
-\hbar)}{\sinh(z-\mu_\ga^{b})}\,.
}
\end{array}
  \eeq
The integer parameters $N_b$ ($N_0=N_n=0$) are the numbers of Bethe
roots $\mu_\be^b$ in the $b$-th group, $b\!=\!1\,,...\,,n\!-\!1$,
$\be\!=\!1\,,...\,,N_b$. They satisfy the system of Bethe equations
(BE): \beq\label{t26}
  \begin{array}{c}
  \displaystyle{
{V_1}\,\prod\limits_{k=1}^N \frac{\sinh(\mu^1_\be-q_k+\hbar )}{\sinh(\mu^1_\be-q_k)}
= {V_{2}}\,\prod\limits_{\ga\neq \be}^{N_{1}}
\frac{\sinh(\mu^1_\be-\mu_\ga^{1}+\hbar
)}{\sinh(\mu^1_\be-\mu_\ga^{1}-\hbar)}
\prod\limits_{\ga=1}^{N_{2}}\frac{\sinh(\mu^1_\be-\mu_\ga^{2}-
\hbar) }{\sinh(\mu^1_\be-\mu_\ga^{2})}}
\\ \\
  \displaystyle{
{V_b}\,
 \prod\limits_{\ga=1}^{N_{b-1}}\frac{\sinh(\mu^b_\be-\mu_\ga^{b-1}+
 \hbar)}{\sinh(\mu^b_\be-\mu_\ga^{b-1})}
 = {V_{b+1}}\,\prod\limits_{\ga\neq \be}^{N_{b}}
\frac{\sinh(\mu^b_\be-\mu_\ga^{b}+\hbar )}{\sinh(\mu^b_\be-\mu_\ga^{b}-\hbar)}
\prod\limits_{\ga=1}^{N_{b+1}}\frac{\sinh(\mu^b_\be-\mu_\ga^{b+1}-
\hbar) }{\sinh(\mu^b_\be-\mu_\ga^{b+1})}}\,.
\end{array}
  \eeq
where $b=2, \ldots , n-1$. In the last
equation it is implied that $N_n=0$.
The BE mean that the eigenvalues
(\ref{t24}) are regular at $z=\mu_{\gamma}^b$.

  It is known that the operators
  \beq\label{mm1}
  \hat M_a = \sum_{j=1}^{N}{\rm e}_{aa}^{(j)}, \quad
 {\rm e}_{aa}^{(j)}=\underbrace{{\sf 1}\otimes \ldots \otimes {\sf 1}}_{j-1}\otimes \,
 {\rm e}_{aa}\otimes \underbrace{{\sf 1}\otimes \ldots \otimes {\sf 1}}_{N-j},
 \eeq
 commute with the transfer matrix. The eigenvectors of the latter, built from
 solutions to the BE, with the number of Bethe roots at level
 $b$ equal to $N_b$, are also eigenvectors of the operators $\hat M_a$
 with the eigenvalues $M_1=N-N_1$, $M_a =N_{a-1}-N_a$, $a=2, \ldots , n$.

The transfer matrix (\ref{t22}) can be represented as a sum over simple poles at
$z=q_k$:
\beq\label{mm2}
\hat T^{\hbox{\tiny{XXZ}}}(z)= \hat C +\sum_{k=1}^N \hat H_{i}^{\hbox{\tiny{XXZ}}}
\coth (z-q_k)
\eeq
(it follows from (\ref{t20}) that it is an $i\pi$-periodic function of $z$).
The coefficients
\beq\label{t27}
  \hat H_{i}^{\hbox{\tiny{XXZ}}}=
  \res\limits_{z=q_i}\hat T^{\hbox{\tiny{XXZ}}}(z)
 \eeq
are quantum (non-local) Hamiltonians of the inhomogeneous spin chain.
They commute with each other: $[\hat H_{i}^{\hbox{\tiny{XXZ}}}, \,
\hat H_{j}^{\hbox{\tiny{XXZ}}}]=0$ and can be simultaneously diagonalized.
This ensures integrability of the model.
 The eigenvalues of the commuting Hamiltonians are given by the formula
 \beq\label{t30}
  \begin{array}{c}
 \displaystyle{
 H_{i}^{\hbox{\tiny{XXZ}}}=V_1\, \sinh \! \hbar \,
\prod\limits_{k\neq i}^N \frac{\sinh(q_i-q_k+\hbar )}{\sinh(q_i-q_k)}
\prod\limits_{\ga=1}^{\,N_{1}}\frac{\sinh(q_i-\mu_\ga^{1}
-\hbar)}{\sinh(q_i-\mu_\ga^{1})}\,,
 }
 \end{array}
  \eeq
where the $\mu_{\gamma}^1$'s are taken from a solution to the BE.
It is easy to see that
\beq\label{mm3}
T^{\hbox{\tiny{XXZ}}}(\pm \infty )= C\pm \sum_{k=1}^{N}H_{k}^{\hbox{\tiny{XXZ}}}=
\sum_{a=1}^n V_a e^{\pm \hbar M_a},
\eeq
hence we get the ``sum rules''
\beq\label{mm4}
C=\sum_{a=1}^n V_a \cosh (\hbar M_a), \qquad
\sum_{k=1}^{N}H_{k}^{\hbox{\tiny{XXZ}}}=\sum_{a=1}^n V_a \sinh (\hbar M_a).
\eeq

\section{Determinant identity}
\setcounter{equation}{0}

Consider a pair of $N\times N$ and $M\times M$ matrices:
 \beq\label{t15}
 \begin{array}{c}
\displaystyle{
 {\mathcal{L}}_{ij}(\{x_i\}_N,\{y_i \}_M,g)=
}
 \\ \ \\
\displaystyle{
 =\,
 \frac{g\sinh\hbar}{\sinh(x_i-x_j+\hbar)}\prod\limits_{k\neq j}^N\frac{\sinh(x_j-x_k+\hbar)}{\sinh(x_j-x_k)}
\prod\limits_{\ga=1}^M\frac{\sinh(x_j-y_\ga)}{\sinh(x_j-y_\ga+\hbar)}\,,
 }
\end{array}
  \eeq
$i\,,j=1\,,...\,,N$ and
 \beq\label{t16}
 \begin{array}{c}
 \displaystyle{
{\widetilde {\mathcal{L}}}_{\al\be}(\{y_i \}_M,\{x_i\}_N,g)=
 }
 \\ \ \\
 \displaystyle{
=\,\frac{g\sinh\hbar}{\sinh(y_\al-y_\be+\hbar)}\prod\limits_{\ga\neq
\be}^M\frac{\sinh(y_\be-y_\ga-\hbar)}{\sinh(y_\be-y_\ga)}
\prod\limits_{k=1}^N\frac{\sinh(y_\be-x_k)}{\sinh(y_\be-x_k-\hbar)}\,,
 }
\end{array}
  \eeq
$\al,\be=1\,,...\,,M$. For definiteness assume that $N\geq M$.
Then the following identity holds true:
 \beq\label{t17}
 \det\limits_{N\times N}
 \Bigl ({\mathcal{L}}\left (\{x_i\}_N,\{y_i \}_M,g\right )
 -\la I\Bigr )\!=\!\det\limits_{(N-M)\times (N-M)}(g S \! -\! \la I)\,
\det\limits_{M\times M}\Bigl ({\widetilde {\mathcal{L}}} \left
(\{y_i \}_M,\{x_i\}_N,g \right )-\la I\Bigr )
  \eeq
Here the matrix $S=S(\hbar )$ (\ref{t09}) entering the r.h.s of (\ref{t17}) is
$(N\!-\!M)\times (N\!-\!M)$ matrix and $I$ is the unity matrix.
The proof of (\ref{t17}) is based
on (\ref{t05}). It is similar to the one given in \cite{GZZ} for the
rational case.

\section{Quantum-Classical duality}
\setcounter{equation}{0}

 \begin{theor}\label{theor1}
Under identification of parameters
 \beq\label{t40}
\eta\nu=\hbar
  \eeq
and
 \beq\label{t41}
 \frac{{\dot q}_j}{\eta}=
 \frac{H_{j}^{\hbox{\tiny{XXZ}}}}{\sinh \hbar}
  \eeq
  where $H_{j}^{\hbox{\tiny{XXZ}}}$ are eigenvalues of the quantum spin chain
  Hamiltonians corresponding to any common eigenstate,
the spectrum of the classical RS Lax matrix (\ref{t04}) is given by
 \beq\label{t42}
 \begin{array}{c}
  \displaystyle{\left.
 \hbox{Spec} \, L^{\hbox{\tiny{RS}}}
 \left (\left \{\dot q_j=\eta \, \frac{H_{j}^{\hbox{\tiny{XXZ}}}}{\sinh \hbar}\right \}_N
 \left \{ q _j\right \} _N, \, \hbar \right )\right |_{BE}}
 \\ \ \\
 \displaystyle{= \Big\{\underbrace{e^{-(N-N_1-1)\hbar}\,V_1,\ldots,
 e^{(N-N_1-1)\hbar}\,V_1}_{N-N_1}\,, \,
 \underbrace{e^{-(N_1-N_2-1)\hbar}\,V_2,\ldots,e^{(N_1-N_2-1)\hbar}
 \,V_2}_{N_1-N_2}\,,\ldots\,,
 }
\end{array}
  \eeq
 $$
 \displaystyle{
 \ldots\,,\underbrace{e^{-(N_{n-2}-N_{n-1}-1)\hbar}\,V_{n-1},
 \ldots,
 e^{(N_{n-2}-N_{n-1}-1)\hbar}\,V_{n-1}}_{N_{n-2}-N_{n-1}}\,,\,
 \underbrace{e^{-(N_{n-1}-1)\hbar}\,V_n,\ldots,e^{(N_{n-1}-1)\hbar}\,V_n}_{N_{n-1}}\Big\}}.
 $$
 \end{theor}

\noindent\underline{\emph{Proof:}}
We can reformulate the statement of the theorem as
\beq\label{Theorem}
\begin{array}{|c|}
  \hline\\ \displaystyle{
\det \left[ L \left (\frac{\eta}{\sinh\hbar}\left\{ H_j^{\hbox{\tiny{XXZ}}}\right\}_N,
 \left\{ q _j\right\}_N, \, \hbar \right )\Bigr|_{BE} - \lambda I\right]=\prod_{a=1}^{n} \det \left[ V_a\, S_{N_a-N_{a-1}}-\lambda I\right]}
\\ \ \\
\hline
\end{array}
\eeq
where $N_0=N$, $N_n=0$, $S_M=S_M(\hbar )$ is the matrix \eqref{t09} of
size $M\!\times\!M$.

The proof of \eqref{Theorem} is performed by successive usage of the
determinant identity (\ref{t17}) and BE (\ref{t26}).
Consider the matrix
\begin{multline}
L^{(0)}_{ij} = L^{\hbox{\tiny{RS}}}_{ij} \left
(\frac{\eta}{\sinh\hbar}\left  \{ H_k^{\hbox{\tiny{XXZ}}}\right
\}_N, \left \{ q _k\right \} _N, \, \hbar \right )   \\
=\frac{V_1\sinh \hbar}{\sinh(q_i-q_j+\hbar)}\prod\limits_{k\neq
j}^N\frac{\sinh(q_j-q_k+\hbar)}{\sinh(q_j-q_k)}
\prod\limits_{\ga=1}^{N_1}\frac{\sinh(q_j-\mu^1_\ga-\hbar)}{\sinh(q_j-\mu^1_\ga)}=\\=
{\mathcal{L}}_{ij}(\{q_k\! -\! \hbar
\}_N,\{\mu_{\gamma}^{\,1}\}_{N_1},V_1)
\end{multline}
\noindent and
 \begin{multline}
L^{(1)}_{\alpha\beta} = \widetilde
{\mathcal{L}}_{\alpha\beta}(\{\mu_{\gamma}^{\,1}\}_{N_1},\{q_i\! -\! \hbar \}_N,V_1) \\
=\frac{V_1\sinh \hbar}{\sinh(\mu_{\alpha}^{1}-\mu_{\beta}^{1}+\hbar)}
\prod\limits_{\ga\neq\beta}^{N_1}\frac{\sinh(\mu^1_\be
-\mu^1_\ga - \hbar)}{\sinh(\mu^1_\be-\mu^1_\ga)}
\prod\limits_{k=1}^N
\frac{\sinh(\mu^1_\beta-q_k+\hbar)}{\sinh(\mu^1_\beta -q_k )}\,,
 \end{multline}
where $\alpha,\beta=1,...,N_1$. Identity \eqref{t17} provides
the relation
 \beq \det\limits_{N\times N} \left( L^{(0)}
-\la I\right)\!=\!\det\limits_{(N-N_1)\times ( N-N_1)}(V_1 S-\la I)\,
\det\limits_{N_1\times N_1}\left(L^{(1)}-\la I\right).
  \eeq
Impose now the BE \eqref{t26}. Then we get
 \beq
L^{(1)}\Bigr|_{BE}=\frac{V_2\sinh\hbar}{\sinh(\mu_{\alpha}^{1}-\mu_{\beta}^{1}+\hbar)}
\prod\limits_{\ga\neq\beta}^{N_1}\frac{\sinh(\mu^1_\be -\mu^1_\ga
+\hbar)}{\sinh(\mu^1_\be-\mu^1_\ga)} \prod\limits_{\ga=1}^{N_2}
\frac{\sinh(\mu_\beta^1-\mu^2_\ga-\hbar)}{\sinh(\mu_\beta^1-\mu^2_\ga)}\,,
 \eeq
\noindent i.e.
 \beq
L^{(1)}\Bigr|_{BE_1}= {\mathcal{L}}_{ij}(\{\mu_\gamma^1\! -\! \hbar
\}_{N_1},\{\mu_{\gamma}^{\,2}\}_{N_2},V_2)\,.
 \eeq
At the next step let us define
 \beq
L^{(2)}_{\alpha\beta} = \widetilde {\mathcal{L}}_{\alpha\beta}( \{\mu_{\gamma}^{\,2}\}_{N_2},\{\mu_{\gamma}^{\,1}-\hbar\}_{N_1},V_2)\,,
\qquad \al,\be=1\,,...\,,N_2\,,
 \eeq
\noindent and again we use \eqref{t17} and \eqref{t26} to get:
 \beq
\det\limits_{N_1\times N_1} \left( L^{(1)}
-\la I \right)\!=\!\det\limits_{N_1-N_2\times N_1-N_2}(V_2 S-\la I)\,
\det\limits_{N_2\times N_2}\left(L^{(2)}-\la I\right)\,.
 \eeq
 \beq
 \left.L^{(2)}\right|_{BE_2}={\mathcal{L}}
 (\{\mu_{\gamma}^{\,2}\!-\!\hbar\}_{N_2},\{
\mu_{\gamma}^{\,3}\}_{N_3},V_3)\,.
\eeq
 $$
\vdots
 $$
\noindent The process of the subsequent  usage of \eqref{t17} and
\eqref{t26} is continued until the last step when equation
\eqref{t26} is used:
 \beq
L^{(n-1)}\Bigr|_{BE_{n-1}}={\mathcal{L}}_{ij}(\{\mu_\gamma^{n-1}\!
-\! \hbar \}_{N_{n-1}}, \{\mu_{\gamma}^{n}\}_0,V_n)\,,
 \eeq
Finally, \eqref{t17} with $N=N_{n-1}, M=0$ yields
 \beq
\det\limits_{N_{n-1}\times N_{n-1}} \left( L^{(n-1)}
-\la I\right)\!=\!\det\limits_{N_{n-1}\times N_{n-1}}(V_n S-\la I)\,.\ \
\ \blacksquare
 \eeq

In order to find the characteristic polynomial of the matrix
$$
L=L^{\hbox{\tiny{RS}}}
 \left (\left \{\dot q_j=\eta \, \frac{H_{j}^{\hbox{\tiny{XXZ}}}}{\sinh \hbar}\right \}_N
 \left \{ q _j\right \} _N, \, \hbar \right )
$$
explicitly, we use the known fact that the coefficient in front of
$\lambda^{N-k}$ in the polynomial $\displaystyle{\det_{N\times N}(\lambda I +A)}$
equals the sum of all diagonal $k\times k$ minors of the matrix $A$. All such minors
can be found using the explicit expression for the determinant
\beq\label{explicit1}
\det_{1\leq i,j\leq k}\frac{\sinh \hbar}{\sinh (q_i\! -\! q_j\! +\! \hbar )}=
\prod_{1\leq i,j\leq k}C(q_i-q_j), \qquad
C(q)=\frac{\sinh ^2 q}{\sinh (q+\hbar)\sinh (q-\hbar)}
\eeq
which can be easily proved or taken from \cite{Ruijsenaars1}.
As a result, we get
$$\displaystyle{\det_{N\times N}(\lambda I +L)=
\sum_{k=0}^{N}J_k \lambda^{N-k}},$$ where
\beq\label{explicit2}
J_k=(\sinh \hbar )^{-k}\!\!\!\!\!\!
\sum_{1\leq i_1< \ldots <i_k \leq N}\!\! H_{i_1}^{\hbox{\tiny{XXZ}}}
\ldots H_{i_n}^{\hbox{\tiny{XXZ}}}\!\!\!\!
\prod_{1\leq \alpha <\beta \leq k}C(q_{i_{\alpha}}-q_{i_{\beta}})
\eeq
Therefore, we have the following system of polynomial equations
for spectrum of the quantum Hamiltonians:
\beq\label{explicit3}
\sum_{1\leq i_1< \ldots <i_k \leq N}H_{i_1}^{\hbox{\tiny{XXZ}}}
\ldots H_{i_k}^{\hbox{\tiny{XXZ}}}\!\!
\prod_{1\leq \alpha <\beta \leq k}C(q_{i_{\alpha}}-q_{i_{\beta}})=
(\sinh \hbar )^k\!\!\!
\sum_{1\leq i_1< \ldots <i_k \leq N}\!\! \lambda_{i_1}\ldots \lambda_{i_k}
\eeq
($k=1, \ldots , N$).
Here $\lambda_i \in \mbox{Spec}\, L$ are given by (\ref{t42}).
Setting $q_i=\hbar x_i$, $H_i^{\hbox{\tiny{XXZ}}}=\hbar \tilde H_i$
and tending $\hbar \to 0$, these equations become the equations
of the universal spectral variety for models of the $XXX$ type
\cite{TsuboiZZ}.

Equations (\ref{explicit3}) at $k=N$ and $k=1$ are easy to check
without directly appealing to the
determinant identity using the side by side products of the BE
and the ``sum rules'' (\ref{mm4}).

At $k=N$ we have the equation
\beq\label{t59a}
\prod_{j=1}^N H_{j}^{\hbox{\tiny{XXZ}}} \, \cdot \!\!\! \prod_{1\leq l<m\leq N}
C(q_l-q_m)=(\sinh \hbar )^N
\prod_{a=1}^n V_a^{M_a}
\eeq
The Bethe ansatz result gives
\beq\label{t59}
\prod\limits_{k=1}^N H_k^{\hbox{\tiny{XXZ}}}=
\left(V_1\sinh\hbar\right)^N
\prod\limits_{i=1}^N\prod\limits_{k\neq i}^N
\frac{\sinh(q_i-q_k+\hbar)}{\sinh(q_i-q_k)}
\prod\limits_{i=1}^N\prod\limits_{\ga=1}^{N_1}
\frac{\sinh(q_i-\mu_\ga^1-\hbar)}{\sinh(q_i-\mu_\ga^1)}\,.
\eeq
The first double product cancels against the product of the
$C$-factors in (\ref{t59a}).
The side by side products of the BE
\beq\label{t60}
  \begin{array}{c}
  \displaystyle{BE_1:\quad
  V_1^{N_1}\prod\limits_{\be=1}^{N_1}\prod\limits_{k=1}^N
  \frac{\sinh(q_k-\mu_\be^1-\hbar)}{\sinh(q_k-\mu_\be^1)}=
  V_2^{N_1}\prod\limits_{\be=1}^{N_1}\prod\limits_{\ga=1}^{N_2}
  \frac{\sinh(\mu_\be^1-\mu_\ga^2-\hbar)}{\sinh(\mu_\be^1-\mu_\be^2)}
  \,,}
 \\ \ \\
  \displaystyle{BE_b:~V_b^{N_b}
  \prod\limits_{\be=1}^{N_b}\prod\limits_{\ga=1}^{N_{b\!-\!1}}
  \frac{\sinh(\mu_\ga^{b-1}-\mu_\be^b-\hbar)}{\sinh(\mu_\ga^{b-1}-\mu_\be^b)}=
  V_{b+1}^{N_b}\prod\limits_{\be=1}^{N_b}
  \prod\limits_{\ga=1}^{N_{b\!+\!1}}
  \frac{\sinh(\mu_\be^b-\mu_\ga^{b+1}-\hbar)}{\sinh(\mu_\be^b-\mu_\ga^{b+1})}
  \
  \,,
  }
  \\ \ \\
  \displaystyle{BE_{n-1}:\quad\quad\quad
  V_{n-1}^{N_{n\!-\!1}}\prod\limits_{\be=1}^{N_{n\!-\!1}}
  \prod\limits_{\ga=1}^{N_{n\!-\!2}}
  \frac{\sinh(\mu_\ga^{n-2}-\mu_\be^{n-1}-\hbar)}{\sinh(\mu_\ga^{n-2}-\mu_\be^{n-1})}
 =V_n^{N_{n\!-\!1}}
 \,,}
\end{array}
  \eeq
form a chain of identities that yields the right hand side of (\ref{t59a}).

At $k=1$ the equation is
\beq\label{explicit4}
\sum_{j=1}^N H_j^{\hbox{\tiny{XXZ}}}=\sinh \hbar \sum_{j=1}^N
\lambda_j , \qquad \lambda_j \in \mbox{Spec}\, L.
\eeq
According to
$$
\sum\limits_{j=0}^{N_{b\!-\! 1}-N_b-1}
e^{-(N_{b\!-\! 1}-N_b-1) + 2j\hbar }=
\frac{\sinh\big(\hbar(N_{b\!-\! 1}-N_b)\big)}{\sinh{\hbar}}
$$
it is exactly the second ``sum rule'' in (\ref{mm4}).

\section{Limiting cases}

\subsection{Limit to the Gaudin-Calogero correspondence}
\setcounter{equation}{0}

{\bf Calogero-Sutherland model.} The Lax matrix of the
Calogero-Sutherland model
 \beq\label{t51a}
 \displaystyle{
L^{\hbox{\tiny{CM}}}_{ij}=\delta_{ij}\, {\dot
q}_j+(1-\delta_{ij})\frac{\nu}{\sinh(q_i-q_j)}
 }
  \eeq
can be represented as
 \beq\label{t511}
 \displaystyle{
L^{\hbox{\tiny{CM}}}=P-\nu D V^{-1}(\epsilon)\p_\epsilon V(\epsilon)
D^{-1}
 }
  \eeq
with matrices $P$, $V$ and $D$ defined in (\ref{t05})-(\ref{t07})
and velocities
 \beq\label{t52}
{\dot q}_j =p_i -\nu\sum\limits_{k\neq i}^N\coth(q_i-q_k)
  \eeq
generated by the Hamiltonian $H^{\hbox{\tiny{CM}}}=\frac12\tr
\left(L^{\hbox{\tiny{CM}}}\right)^2$.
The representation (\ref{t511}) follows from (\ref{t05}) in the
non-relativistic limit $\eta\rightarrow 0$:
 \beq\label{t53}
 \displaystyle{
L^{\hbox{\tiny{RS}}}={\sf 1}_{N\times N}+\eta\, L^{\hbox{\tiny{CM}}}+
O(\eta^2)\,.
 }
  \eeq
  It follows from (\ref{t06}) and (\ref{t09}) that
 \beq\label{t54}
 \displaystyle{
\p_\epsilon V(\epsilon)=-\log S(\nu) V\,,
 }
  \eeq
where
 \beq\label{t55}
 \displaystyle{
-\log S_{ij}(\zeta)=\delta_{ij}\left( (2i-1-N)\zeta \right)\,.
 }
  \eeq
Therefore,
 \beq\label{t56}
 \displaystyle{
{\rm
Spec}(L^{\hbox{\tiny{CM}}}\left.\right|_{P=0})=\left\{{-(N-1)\nu},
{-(N-3)\nu},...,{(N-3)\nu},{(N-1)\nu}\right\}\,.
 }
  \eeq

\vskip3mm

\noindent {\bf The trigonometric Gaudin model} appears in the limit
 $\varepsilon \to 0$  from the inhomogeneous $XXZ$ spin chain
with the transfer matrix $\hat T^{\hbox{\tiny{XXZ}}}(z; \{q_i\},
V^{\varepsilon}, \varepsilon \hbar )$, where
 $$V^\epsilon={\sf 1}_{n\times n}+\varepsilon\,
 \hbox{diag}(v_1,...,v_n)+O(\varepsilon^2)\,.$$
The expansion as $\varepsilon\to 0$,
 \beq\label{expansion}
 \hat T^{\hbox{\tiny{XXZ}}}(z;\{q_i\}, V^{\varepsilon}, \varepsilon \hbar )=
 nI + \varepsilon \hat T_1(z;\{q_i\}) + \varepsilon^2 \hat T_2(z;\{q_i\}) + O(\varepsilon^3),
\eeq
$$
\hat T_1 (z; \{q_i\})=\mbox{tr}\, v \, I+\hbar \sum_i C_1^{(i)}\coth (z-q_i),
\qquad C_1^{(i)}=\sum_a {\rm e}_{aa}^{(i)},
$$
 defines the commuting Gaudin Hamiltonians
\beq
  \hat H^{\hbox{\tiny{G}}}_{\, i}=
  \res\limits_{z=q_i}\hat T_2(z;\{q_i\})\,,
\eeq
 \beq\label{HG}
\hat H^{\hbox{\tiny{G}}}_{\, i}= \sum_a v_a {\rm e}_{aa}^{(i)}+
\sum_{j\neq i} \frac{\hbar}{\sinh(q_i-q_j)}
\left(  \sum_{a\ne b} {\rm e}_{ab}^{(i)}{\rm e}_{ba}^{(j)} +
\cosh(q_i-q_j)  \sum_{a} {\rm e}_{aa}^{(i)}{\rm e}_{aa}^{(j)} \right).
 \eeq
 The commutativity of the Gaudin Hamiltonians follows from
 commutativity of the transfer matrices, taken into account that
 the term $\hat T_1 (z; \{q_i\})$ is central.
Their eigenvalues can be found using (\ref{t30}) and tending
$\varepsilon\rightarrow 0$. This gives
 \beq\label{HGeig}
\displaystyle{
H_i^{\hbox{\tiny{G}}}=v_1+\hbar\sum\limits_{k\neq i}^N \coth(q_i-q_k)
%
%
-\hbar\sum\limits_{\ga=1}^{N_{1}} \coth ({q_i-\mu_\ga^{1}})}
\eeq
with the BE at level $b$ of the form
 \beq\label{BEG}
 \begin{array}{c}
 {v_b}+\delta_{1b} \hbar\sum\limits_{k=1}^N
\coth({\mu^b_\be-q_k})
+\hbar\sum\limits_{\ga=1}^{N_{b\!-\!1}}\coth(\mu^b_\be-\mu_\ga^{b\!-\!1})=
\\
={v_{b\!+\!1}}+2\hbar\sum\limits_{\ga\neq \be}^{N_{b}}\coth({\mu^b_\be-\mu_\ga^{b}})
 -\hbar\sum\limits_{\ga=1}^{N_{b\!+\!1}}\coth({\mu^b_\be-\mu_\ga^{b\!+\!1}})
 \end{array}
 \eeq
 where $b\!=\!1\,,...\,,n\!-\!1$, $N_0 \! =\! N_{n}\!=\!0$,
$\be\!=\!1\,,...\,,N_b$. The matrix $v=\mbox{diag}(v_1\,,...\,,v_n)$
is the twist matrix of the Gaudin model.
Similarly to the  (XXZ) spin chain case we use the notation
$H^{\hbox{\tiny{G}}}_i (\{q_i\}_N , \{ \mu_{\alpha}^1\}_{N_1})$ for
the function given by the r.h.s. of (\ref{HGeig}). When the set $\{
\mu_{\alpha}^1\}_{N_1}$ is taken from a solution of the system of BE
(\ref{BEG}) this function is equal to some eigenvalue of the
Hamiltonian.

\noindent {\bf Determinant identity.} 
Introduce the following pair of matrices
 \beq
\begin{array}{c}
 \displaystyle{
\mathcal{L}_{ij}(\{x_i\}_N,\{y_i\}_M,\omega)
 }
 \\ \ \\
 \displaystyle{
=\delta_{ij}\left(\om+\sum\limits^N_{k\neq
i}\nu \coth (x_i-x_k)+\sum\limits_{\ga=1}^M\nu \coth (y_\ga-x_i)
\right)+(1-\delta_{ij})\frac{\nu}{\sinh (x_i-x_j)\,,} }
\end{array}
  \eeq
where $i\,,j=1\,,...\,,N$ and
 \beq
\begin{array}{c}
 \displaystyle{
{\widetilde {\mathcal{L}}}_{\al\be}(\{y_i\}_M,\{x_i\}_N,\omega)
 }
 \\ \ \\
 \displaystyle{
=\delta_{\al\be}\left(
\om-\sum\limits^M_{\ga\neq\al}\nu \coth(y_\al\!-\!y_\ga)
-\sum\limits^N_{k=1}\nu \coth (x_k\!-\!y_\al)\right)+
\left(1-\delta_{\al\be}\right) \frac{\nu}{\sinh (y_\al\!-\!y_\be)}\,,
}
\end{array}
  \eeq
\noindent where $\al,\be=1\,,...\,,M$. The relation between their
determinants is given as follows:
 \beq\label{di2}
\begin{array}{c}
 \det\limits_{N\times N}
 \Bigl ({\mathcal{L}}\left (\{x_i\}_N,\{y_i \}_M,\om\right )
 -\la I\Bigr )
 \\ \ \\
 =\det\limits_{(N-M)\times (N-M)}(\om I+ \log S - \la I)\,
\det\limits_{M\times M}\Bigl ({\widetilde {\mathcal{L}}} \left
(\{y_i \}_M,\{x_i\}_N,g \right )-\la I\Bigr )\,,
\end{array}
  \eeq
where $\log S=\log S(\nu)$ (\ref{t55}) entering r.h.s of (\ref{di2})
is the $(N\!-\!M)\times (N\!-\!M)$ diagonal matrix.

\vskip3mm

\noindent {\bf Quantum-classical duality} between the classical
Calogero-Sutherland system and the quantum Gaudin model is given by the
following statement:
 \begin{theor}\label{theor2}
Under identification of the parameters
 \beq
\nu=\hbar
  \eeq
and
 \beq
 {\dot q}_j=
 \frac{1}{\hbar}\,H^{\hbox{\tiny{G}}}_j (\{q_i\}_N , \{ \mu_{\alpha}^1\}_{N_1})
  \eeq
  where $H^{\hbox{\tiny{G}}}_j$ are eigenvalues of the quantum Gaudin
  Hamiltonians corresponding to any common eigenstate,
the spectrum of the Lax matrix (\ref{t51a}) is equal to
 \beq
 \begin{array}{c}
  \displaystyle{
 \hbox{Spec} \, L^{\hbox{\tiny{CM}}}
 \left (\frac{1}{\hbar}\left  \{ H_j^{\hbox{\tiny{G}}}\right \}_N,
 \left \{ q _j\right \} _N, \, \hbar \right )\Bigr |_{BE}}=
 \\ \ \\
 \displaystyle{=
 \Big\{\underbrace{v_1\! -\! (N\! -\! N_1\! -\! 1)
 \hbar, \, \ldots,\, v_1 \! +\! (N\! -\! N_1\! -\! 1)\hbar}_{N\! -\! N_1}\,,}
  \\ \ \\
 \displaystyle{\underbrace{v_2\! -\! (N_1\! -\! N_2\! -\! 1)\hbar,\, \ldots, \,
 v_2\! +\! (N_1\! -\! N_2\! -\! 1)\hbar}_{N_1\! -\! N_2}\,,
 }
\end{array}
\eeq
 $$
 \displaystyle{
 \ldots\,,
 \underbrace{v_n\! -\! (N_{n-1}\! -\! 1)\hbar,\, \ldots, \,
 v_n\! +\! (N_{n-1}\! -\! 1)\hbar}_{N_{n-1}}\Big\}}.
 $$
 \end{theor}
The proof Theorem \ref{theor2} is similar to Theorem
\ref{theor1}. Similarly to the non-degenerate $XXZ$ case, the
eigenvalues of the Lax matrix form ``strings'' centered at the $v_a$'s.
The distance between to subsequent eigenvalues in any string is $2\hbar$.

\subsection{Limit to XX model}
\setcounter{equation}{0}

The XXZ model has a limit $\hbar\rightarrow i\pi/2$ called the XX model.
 The latter is often referred to as the free-fermion model, which is due to the
 fact that the XX Hamiltonian may be mapped to a creation-annihilation form that
 corresponds to a system of non-interacting fermions on the 1D lattice.
As none of the $R-$matrix entries vanish
at $\hbar =i\pi/2$, the eigenvalues of the transfer matrix simplify insignificantly:
 \beq\label{t44}
  \begin{array}{c}
 \displaystyle{
T^{\hbox{\tiny{XX}}}_{}(z)= i^{N-N_1}\,V_1\,
\prod\limits_{k=1}^N\coth(z-q_k)
\prod\limits_{\ga=1}^{\,N_1}\coth(z-\mu_\ga^1)~+
\quad\quad\quad
 }
 \\ \ \\
 \displaystyle{\quad\quad\quad+
\sum\limits_{b=2}^n i^{N_{b-1}-N_b}\,V_b \,
\prod\limits_{\ga=1}^{N_{b\!-\!1}}\coth(z-\mu_\ga^{b\!-\!1})
\prod\limits_{\ga=1}^{\,N_{b}}\coth(z-\mu_\ga^b).
 }
 \end{array}
  \eeq
So do the eigenvalues of the quantum Hamiltonians:
 \beq\label{t45}
  \begin{array}{c}
 \displaystyle{
 H_{j}^{\hbox{\tiny{XX}}}=i^{N-N_1}\, V_1\,
\prod\limits_{k\neq j}^N \coth(q_j-q_k)
\prod\limits_{\ga=1}^{\,N_{1}} \coth(q_j-\mu_\ga^{1})\,.
 }
 \end{array}
  \eeq
What is special about the free-fermion point is the simplification of the
BE (\ref{t26}) due to
collapse of one of the two products in the right hand sides caused by periodicity of
the $\sinh$-function along the imaginary axis:
 \beq\label{t46}
  \begin{array}{c}
  \displaystyle{BE_1^{}:\quad
  i^N\,V_1\,\prod\limits_{k=1}^N \coth(\mu^1_\be-q_k)
 =V_2\,(-1)^{N_1-1}\,i^{-N_2}
 \prod\limits_{\ga=1}^{N_2}\coth(\mu^1_\be-\mu_\ga^2)\,,}
 \\ \ \\
  \displaystyle{BE_b^{}:~
  i^{N_{b\!-\!1}}V_b\,\prod\limits_{\ga=1}^{N_{b\!-\!1}}
  \coth(\mu^b_\be-\mu_\ga^{b-1})=V_{b+1}\,(-1)^{N_b-1}i^{-N_{b\!+\!1}}
 \prod\limits_{\ga=1}^{N_{b\!+\!1}}\coth(\mu^b_\be-\mu_\ga^{b+1})\,,
  }
  \\ \ \\
  \displaystyle{BE_{n-1}^{}:\quad\quad\quad
  i^{N_{n\!-\!2}}\,V_{n-1}\,\prod\limits_{\ga=1}^{N_{n\!-\!2}}
  \coth(\mu^{n-1}_\be-\mu_\ga^{n-2})=
 V_n\,(-1)^{N_{n\!-\!1}-1}
 \,,}
\end{array}
  \eeq
where $b=2,\dots,n-2.$

 The equations for the spectrum (\ref{explicit3}) acquire the form
 \beq\label{explicit3a}
\sum_{1\leq i_1< \ldots <i_k \leq N}H_{i_1}^{\hbox{\tiny{XX}}}
\ldots H_{i_k}^{\hbox{\tiny{XX}}}\!\!
\prod_{1\leq \alpha <\beta \leq k}\tanh ^2 (q_{i_{\alpha}}-q_{i_{\beta}})=
i^k\!\!\! \!\!\!\!
\sum_{1\leq i_1< \ldots <i_k \leq N}\!\! \lambda_{i_1}\ldots \lambda_{i_k}
\eeq
($k=1, \ldots , N$). The eigenvalues of the Lax matrix are
$i^{-(M_a-1)}(-1)^{\alpha}V_a$, $a=1, \ldots n$, $\alpha =0,1, \ldots M_a-1$.

\begin{small}

\end{small}

\end{document}